\documentclass{elsart}

\usepackage{graphicx}

\usepackage{amssymb}

\begin{document}

\begin{frontmatter}




\title{Long-range correlations and nonstationarity 
in the Brazilian stock market\thanksref{salinas}}
\thanks[salinas]{This article is dedicated to Prof.~S.~R.~Salinas 
in his 60th birthday.}

\author{R. L. Costa}\and\author{G. L. Vasconcelos\corauthref{glv}}
\corauth[glv]{Corresponding author. Tel.: +55-81-3271-8450; 
fax: +55-81-3271-0359}
\ead{giovani@lftc.ufpe.br}
\address{
Laborat\'orio de F\'{\i}sica Te\'orica e Computacional,
Departamento de F\'{\i}sica, Universidade Federal de Pernambuco,
50670-901, Recife, Brazil.
}

\begin{abstract}
We report an empirical study of the Ibovespa index of the S\~ao Paulo
Stock Exchange in which we detect the existence of long-range
correlations.  To analyze our data we introduce a rescaled variant of
the usual Detrended Fluctuation Analysis that allows us to obtain the
Hurst exponent through a one-parameter fitting.  We also compute a
time-dependent Hurst exponent $H(t)$ using three-year moving time
windows.  In particular, we find that before the launch of the Collor
Plan in 1990 the curve $H(t)$ remains, in general, well above 1/2, while
afterwards it stays close to 1/2.  We thus argue that the structural
reforms set off by the Collor Plan has lead to a more efficient stock
market in Brazil. We also suggest that the time dependence of the
Ibovespa Hurst exponent could be described in terms of a
multifractional Brownian motion.

\end{abstract}

\end{frontmatter}

\section{Introduction}

In recent years, it has been realized that many problems from
Economics can be studied with the standard ``toolkit'' of statistical
physics \cite{stanley0,potters}.  For instance, asset prices in
financial markets are commonly described in terms of a geometric
Brownian motion, an assumption that is at the heart of the so-called
{\it Efficient Market Hypothesis} (EMH). Thus, in the EMH scenario the
returns on a given stock follow an uncorrelated Gaussian process
(white noise).  In more common parlance, the EMH says that all
information available about a given financial asset is already
reflected on its current price, so that knowing the asset past history
does not in any way help us to predict future prices. Although the EMH
stands as a cornerstone of modern Finance \cite{ingersoll}, deviations
from efficiency have been recently observed in many different
financial markets \cite{stanley0,potters}. In such `inefficient'
markets the empirical data violate either the independence or the
Gaussian assumptions of the EMH. The former instance of `inefficiency'
is of particular interest to us here because it implies  the 
existence of long-range correlations that are not accounted for in the
standard EMH model.

A time series can be tested for correlation in many different ways
\cite{taqqu}. One general methodology consists in
estimating how a certain fluctuation measure, to be denoted here
generically by $F$, scales with the size $\tau$ of the time window
considered. Specific methods, such as the Hurst rescaled range
analysis \cite{hurst} or the Detrendend Fluctuation Analysis
\cite{stanley1,jafferson}, differ basically on the choice of the fluctuation
measure; see below for more details about these two methods.  If the
time series is uncorrelated one expects that $F\sim\tau^{1/2}$, as is
the case for the standard Brownian motion. On the other hand, if
$F\sim\tau^H$ with $H\ne1/2$ one then says that the time series has
long-term memory, with $H>1/2$ ($H<1/2$) meaning
persistence (antipersistence). The exponent $H$ is generally
referred to as the Hurst exponent.

In the present paper we analyze the behavior of the S\~ao Paulo Stock
Exchange Ibovespa index, which is the main index of the Brazilian
stock market. We have carried out a Detrended Fluctuation Analysis
(DFA) of the Ibovespa covering over 30 years of data, since its
inception in 1968 until the year 2001. Here we introduce however a
rescaled variant of the original DFA \cite{stanley1} that has the
advantage of allowing us to determine the Hurst exponent $H$ with a
one-parameter fitting. For the complete Ibovespa time series we find
$H=0.6$, indicating that the Ibovespa index exhibits persistence. We
have also computed a `local' Hurst exponent $H(t)$ in moving
three-year time windows and found that $H(t)$ varies considerably over
time.  In particular, we find that the curve $H(t)$ undergoes a
distinct change in character around the year 1990. More specifically,
we observe that before 1990 the exponent $H(t)$ is always greater than
1/2. Then around the year 1990 $H(t)$ drops quite rapidly towards 1/2
and stays  (within some fluctuation) around this value afterwards. We
identify this drastic change in $H(t)$ as a consequence of the
economic plan adopted in March 1990, the so-called Collor Plan, that
marked the beginning of structural reforms in the Brazilian economy.
This interpretation is confirmed by separate analyses of the Ibovespa
returns prior and after the Collor Plan, where in the former case we
find $H=0.63$ while for the latter period we have $H\approx0.5$.  We
thus argue that the process of opening and modernization of the
economy started by the Collor Plan has led to a more efficient stock
market in Brazil, in the sense that the Hurst exponent $H$ for this
period is close to the value predicted by the EMH.

The fact that its Hurst exponent changes over time indicates that the
Ibovespa follows a multifractal (rather than monofractal) process. We
thus suggest that the Ibovespa basic dynamics could perhaps be
captured by the so-called multifractional Brownian motion (MFBM)
\cite{mfbm}, which is a generalized fractional Brownian motion with a
time-dependent Hurst exponent $H(t)$. One interesting property of the
MFBM is that it is a Gaussian {\it additive} process, while the
multifractal processes often considered in physics \cite{feder,turcotte} and
economics \cite{mandelbrot} are in general multiplicative.

The paper is organized as follows. In Sec.~2 we describe our data. In
Sec.~3 we present some background material about fractional Brownian
motion and give a brief description of the Detrended Fluctuation
Analysis.  In particular, we discuss a rescaled version of the DFA
that allows us to determine the Hurst exponent $H$ through a
one-parameter fitting.  The results of this rescaled DFA for the
Ibovespa time series are shown in Sec.~4. In Sec.~5 we offer a succinct
discussion of the multifractional Brownian motion and its possible
relevance to financial time series. In Sec.~6 we review our main
results and conclusions. For completeness, we present in Appendix A
the results of the Hurst analysis for our Ibovespa time series.  It is
shown there that in this case the Hurst method is slightly less
reliable than the DFA in detecting long-correlation, thus justifying
our choice of restricting ourselves to the DFA in the main portion of
the present paper.

\section{The Data}

\begin{figure}[b]
\begin{center}
\includegraphics*[width=.7\columnwidth]{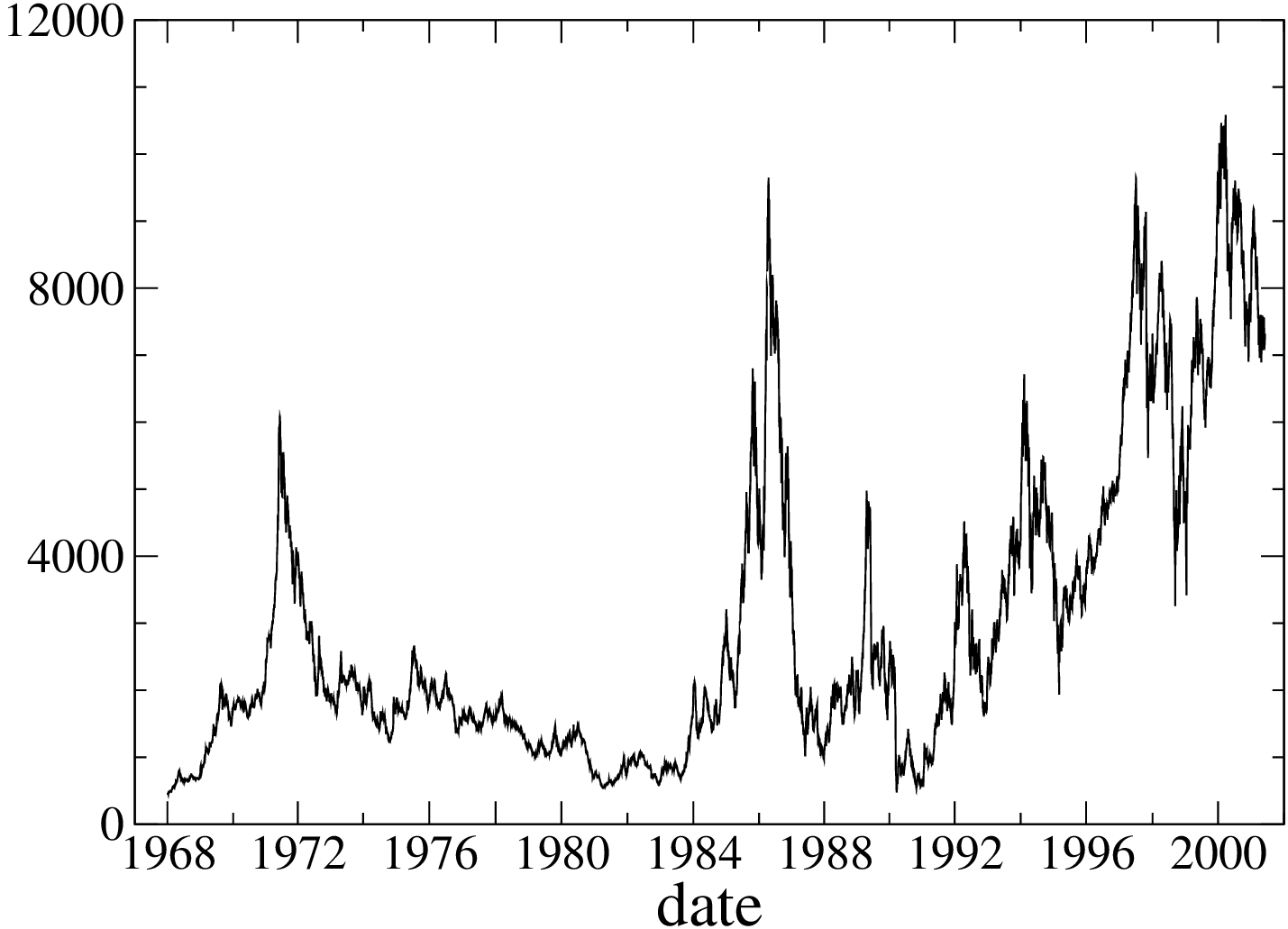}
\caption{Daily closing values of the deflated Ibovespa index 
in the period January 1968--May 2001.} 
\label{fig:ibovespa}
\end{center}
\end{figure}

\begin{figure}
\begin{center}
\includegraphics*[width=.7\columnwidth]{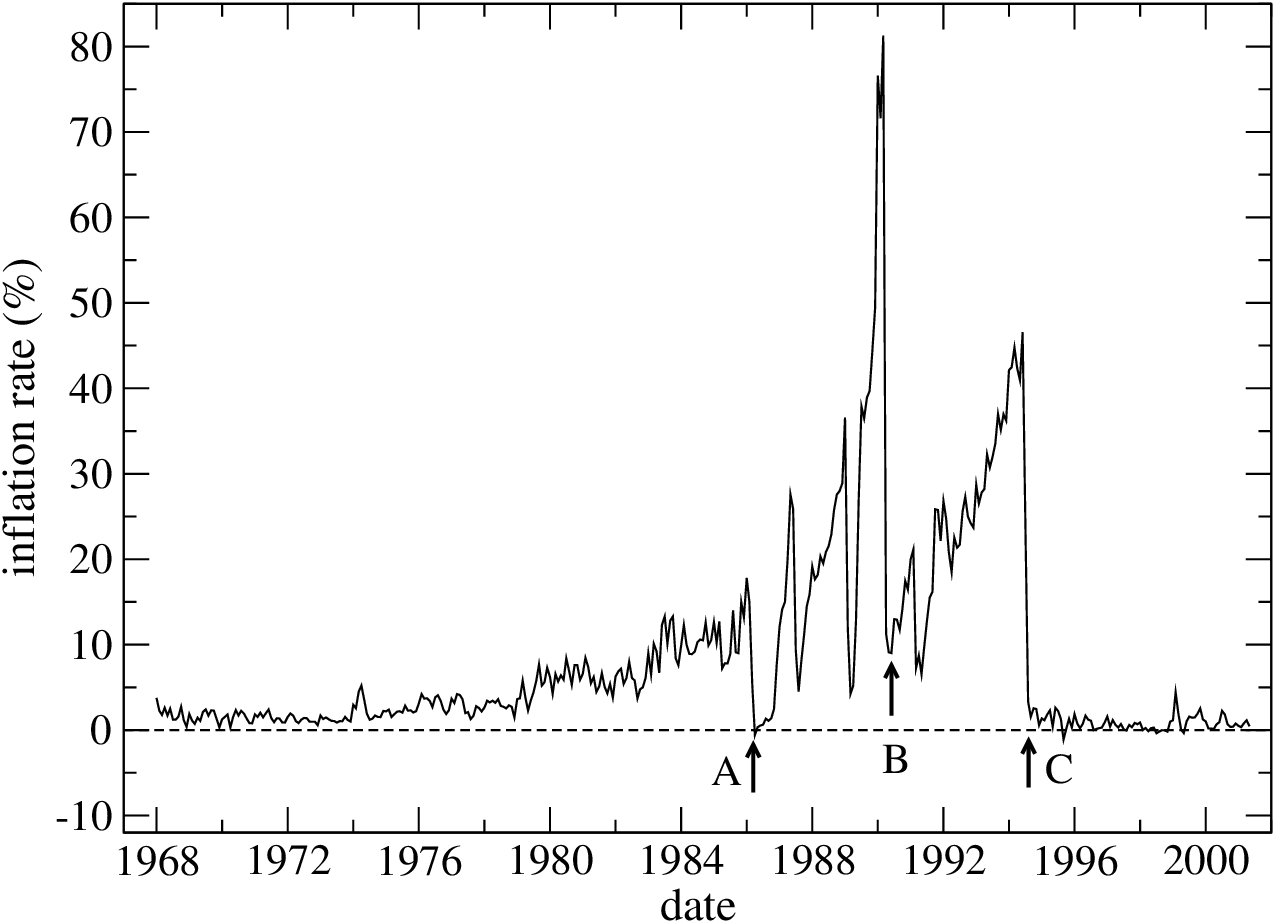}
\caption{Brazilian monthly inflation rate in the period January 
1968--May 2001. The arrows indicate the time of launch of the
following economic plans: Cruzado Plan (A), Collor Plan (B), and the
Real Plan (C); see text.}
\label{fig:igp}
\end{center}
\end{figure}

The Ibovespa index represents the present value of a self-financing
hypothetical portfolio made up of the most traded stocks on the S\~ao
Paulo Stock Exchange.  In this paper we have analyzed the daily
closing values of the Ibovespa since its inception (January 02, 1968)
until recent times (May 31, 2001), amounting to 8209 trading days. The
nominal figures of the daily Ibovespa were deflated by Brazil's
general price index (IGP-DI, in its Portuguese acronym) and converted
to Brazilian Reais (BRL) in values of August 1994. The corresponding
time series for the deflated Ibovespa is shown in
Fig.~\ref{fig:ibovespa}, while the IGP-DI price index is plotted in
Fig.~\ref{fig:igp} in monthly rates.

In the period comprised in our study many economic events took place
in Brazil and abroad that had a direct impact on the Brazilian stock
market. For the benefit of the reader less familiar with Brazil's
recent economic history, it was thought desirable to recall here some
of the main events whose effects can be clearly distinguished in the
data shown in Figs.~\ref{fig:ibovespa} and \ref{fig:igp}.  For
instance, the first large peak in the Ibovespa in the early 1970's
corresponds to the so-called `Brazilian economic miracle' (during the
military regime that took power in 1964 and ended in 1985) when
Brazil's GDP grew over 10\% a year. After the first oil crises in 1973
the Brazilian economy entered a long period of slower growth and
steadily increasing inflation, as reflected in the consistent decline
of the Ibovespa from 1973 to 1984 (Fig.~\ref{fig:ibovespa}) and a
corresponding surge of the inflation rate (Fig.~\ref{fig:igp}).  From
1986 to 1991 six largely unsuccessful economic plans were adopted by
the succeeding governments in attempts to control inflation.  The
effects of these economic plans are very clearly distinguished in
Fig.~\ref{fig:igp} as they correspond to sudden drops in the inflation
rate.  For example, the first of such plans, the so-called Cruzado
Plan, was launched in February 1986 and led not only to a sharp
decline in the inflation rate (arrow A in Fig.~\ref{fig:igp}) but also
to a strong peak in the Ibovespa index (see Fig.~\ref{fig:ibovespa}).
Shortly afterwards, however, inflation picks up again and the Ibovespa
recedes to pre-Cruzado levels. Although it also failed to rein in
inflation, a plan of more lasting impact on the Brazilian economy was
the Collor Plan (arrow B in Fig.~\ref{fig:igp}), as we shall see
later. It should also be recalled that the Real Plan, launched in July
1994 (arrow C in Fig.~\ref{fig:igp}), has so far succeeded in
stabilizing the economy and keeping inflation at bay.

\begin{figure}[t]
\begin{center}
\includegraphics*[width=.7\columnwidth]{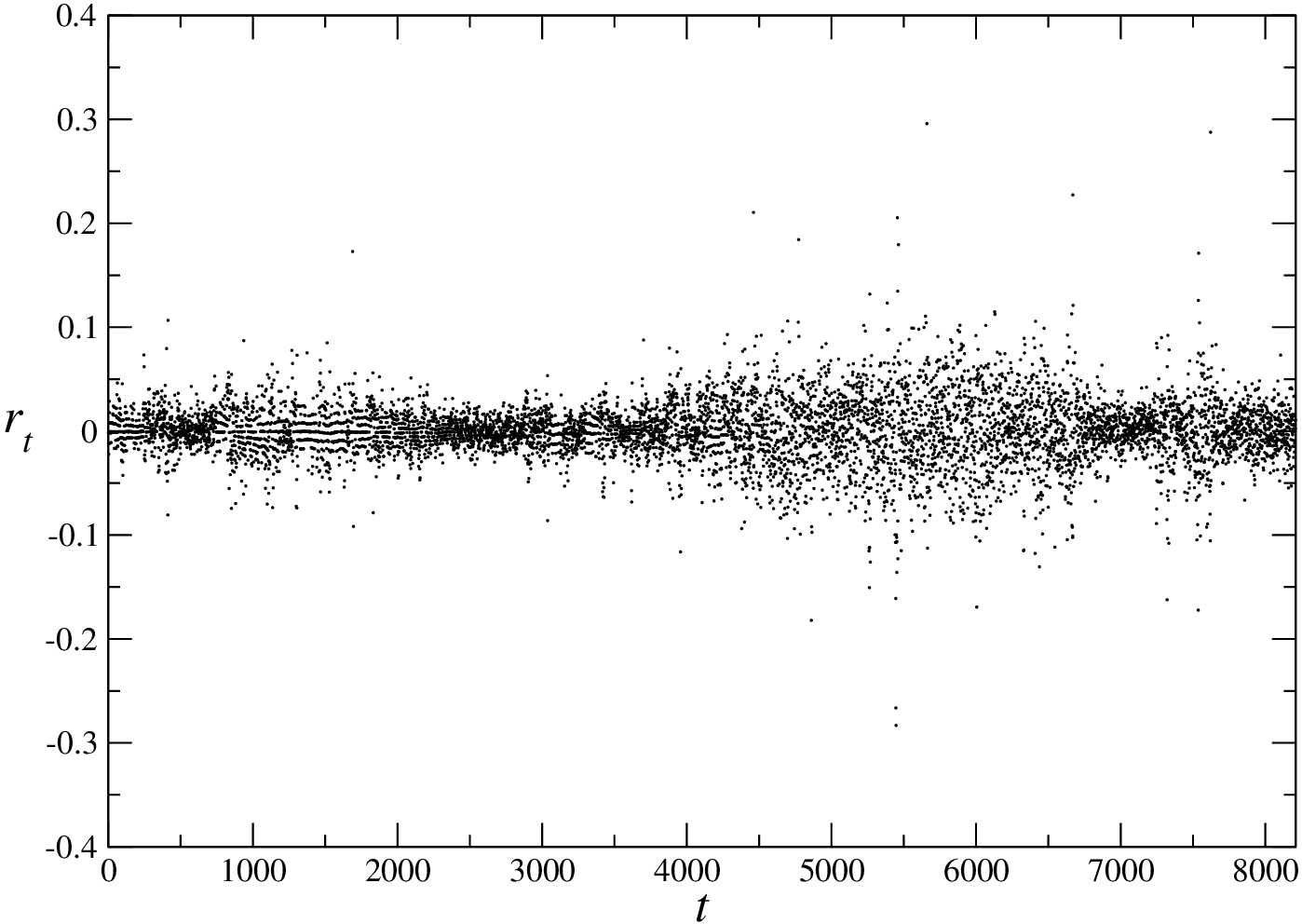}
\caption{Daily returns of the Ibovespa in the period January 
02, 1968 ($t=0$) through May 30, 2001 ($t=8207$).} 
\label{fig:returns}
\end{center}
\end{figure}

In what follows we shall carry out a detailed analysis of the Ibovespa
daily returns, which we define in the usual way.  Let $y_t$ denote the
closing value of the Ibovespa at time $t$, where $t$ is
measured in trading days and $t=0$ corresponds to the date January 02,
1968, when the Ibovespa was created.  The return $r_t$ at time $t$ is
given by the logarithmic difference between consecutive daily
values:
\begin{equation}
r_t\equiv \ln y_{t+1}-\ln y_{t}= \ln \left( \frac{y_{t+1}}{y_{t}}\right).
\end{equation}
In Fig.~\ref{fig:returns} we plot the daily returns $r_t$ for the
Ibovespa index.  In what follows we wish to investigate the existence
of long-term dependence in the Ibovespa daily returns. Before going
into this analysis, however, we shall first present some background
material about correlated time series and describe briefly the method
we will use to detect correlation in our data.

\section{Correlated Time Series}

One of the simplest and most important stochastic models that display
long-term dependence is the fractional Brownian motion (FBM), whose
main properties we will briefly review below. Following this, we shall
then describe the Detrended Fluctuation Analysis (DFA) which, as
already noted, is the main method we will use to estimate the
Hurst exponent $H$ for our empirical data.  (Several other estimators
for $H$ have been discussed in the literature; see, e.g.,
Ref.~\cite{taqqu} for a comparison among some of them.)

\subsection{Fractional Brownian motion}

We recall here that the fractional Brownian motion $\{B_H(t), t>0\}$ is a
Gaussian process with zero mean and stationary increments whose
variance and covariance are given by
\begin{eqnarray}
&&E[B_H^2(t)]=\sigma^2t^{2H}, \label{eq:varBH}\\
&&E[B_H(s)B_H(t)]=\frac{1}{2}\sigma^2\left(s^{2H}+
t^{2H}-|t-s|^{2H}\right), \label{eq:corBH}
\end{eqnarray}
where $0<H<1$,  $\sigma>0$, and $E[\cdot]$ denotes expected value.
The process $B_H(t)$ is statistically self-similar (or more exactly
self-affine) in the sense of finite-dimensional distributions:
\begin{equation}
B_H(at)\stackrel{d}{=}a^HB_H(t),
\label{eq:aH}
\end{equation}
for all $a>0$, where $\stackrel{d}{=}$ means equality in distribution.  

The parameter $H$ is called the self-similarity exponent or the Hurst
exponent. For $H=1/2$ the process $B_H(t)$ corresponds to the standard
Brownian motion, in which case the increments $X_t=B_H(t+1)-B_H(t)$
are statistically independent and represent the usual white
noise. On the other hand, for $H\ne1/2$ the increments $X_t$,
known as fractional white noise, display long-range correlation in the
sense that
\begin{equation}
E[X_{t+h}X_t]\approx\sigma^22H(2H-1)h^{2H-2} \quad \mbox{for} \quad h\to\infty,
\end{equation}
as one can easily verify from (\ref{eq:varBH}) and (\ref{eq:corBH}).
Thus, if $1/2<H<1$ the increments of the FBM are positively correlated
and we say that the process $B_H(t)$ exhibits persistence. Likewise,
for $0<H<1/2$ the increments are negatively correlated and the FBM is
said to show antipersistence.

\subsection{Detrended fluctuation analysis}
 
The Detrended Fluctuation Analysis (DFA), which is a modification of
the usual variance analysis, was proposed independently albeit with
different names in Refs.~\cite{stanley1} and \cite{jafferson} . (In
Ref.~\cite{jafferson} the analogous of the fluctuation function
$F(\tau)$ defined below was termed Straight Line Roughness.)  The DFA
has the advantage over the standard variance analysis of being able to
detect long-term dependence in nonstationary time series
\cite{stanley1}.  The idea of the method is to subtract possible
deterministic trends from the original time series and then analyze
the fluctuation of the detrended data, as explained below.

To implement the DFA, we first integrate the original time series
$\{r_t\}_{t=1,...,T}$ to obtain the cumulative time series $X(t)$:
\begin{equation}
X(t)=\sum_{t^\prime=1}^{t}(r_{t^\prime}-\overline{r}), \quad t=1,...,T,
\end{equation}
where
\begin{equation}
\overline{r}=\frac{1}{T}\sum_{t'=1}^{T} r_{t'} .
\end{equation}
Next we break up $\{X(t)\}$ into $N$ non-overlapping time intervals,
$I_n$, of equal size $\tau$, where $n=0,1,...,N-1$ and $N$ corresponds
to the integer part of $T/\tau$. We then  introduce the local trend
function $Y_\tau(t)$ defined by
\begin{equation}
Y_\tau (t)=a_n+b_nt \quad \mbox{for} \quad t\in I_n,
\end{equation}
where the coefficients $a_n$ and $b_n$ represent the least-square linear
fit of $X(t)$ in the interval $I_n$. We now compute the
fluctuation function $F(\tau)$ defined as the root mean deviation of
$X(t)$ with respect to the trend function $Y_\tau(t)$:
\begin{equation}
F(\tau)=\sqrt{\frac{1}{T }\sum _{t=1}^{T }\left[X(t)-Y_\tau(t)\right]^2}.
\label{eq:F}
\end{equation}
In the original DFA \cite{stanley1} the Hurst exponent $H$ is obtained
directly from the scaling behavior of the above fluctuation function:
$F(\tau)\sim\tau^H$. For reasons that will become apparent shortly, we
find it convenient to introduce a rescaled (dimensionless) fluctuation
function $F_S(\tau)$,
\begin{equation}
F_S(\tau)\equiv\frac{F(\tau)}{S},
\label{eq:FS}
\end{equation} 
where $S$ is the data standard deviation
\begin{equation}
S=\sqrt{\frac{1}{T}\sum _{t=1}^T\left(
r_{t}-\overline{r}\right)^{2}}.
\end{equation}
The rescaled fluctuation function $F_S(\tau)$ obviously has the same
functional form as the original $F(\tau)$ and so
we write
\begin{equation}
F_S(\tau) = C \tau^H, 
\label{eq:alpha}
\end{equation}
where $C$ is a constant independent of the time lag $\tau$. From
here on, we shall drop the subscript from $F_S(\tau)$, with the
understanding that we will be working solely with the rescaled
fluctuation function.

In a double-logarithmic plot the relationship (\ref{eq:alpha})
yields a straight line whose slope is precisely the exponent $H$, 
and so the empirical value for $H$ can  be easily obtained from a linear
regression (in log-log scale) of the corresponding data for $F(\tau)$. One
practical problem with this method, however, is that one needs to
choose an appropriate interval within which to perform the linear fit.
For instance, the first few points at the low end of the graph of
$F(\tau)$ should be disregarded because in this region the detrending
procedure removes too much of the fluctuation---this effect accounts
for the usual `bending down' of $F(\tau)$ for small $\tau$. For large
$\tau$, on the other hand, there are few boxes $I_n$ for a proper
averaging to be made and hence the values of $F(\tau)$ are not
statistically reliable in this case. Furthermore, it has been found
\cite{stanley2} that the values obtained for $H$ using this 
procedure are somewhat dependent on the choice of the fitting interval.  In
order to avoid some of these difficulties we shall use here a
slightly different procedure to estimate the exponent $H$ from the
graph of $F(\tau)$.

To do this, we shall rely on the fact that for the fractional Brownian
motion, i.e., for $X(t)=B_H(t)$, the (rescaled)
fluctuation function $F(\tau)$ can be computed exactly \cite{taqqu}:
\begin{equation}
F_H(\tau) = C_H \tau^H, 
\label{eq:FH}
\end{equation}
where
\begin{equation}
C_H=\frac{2}{2H+1}+\frac{1}{H+2}-\frac{2}{H+1}.
\label{eq:CH}
\end{equation}
In (\ref{eq:FH}) we have added a subscript $H$ to the fluctuation
function $F(\tau)$ to denote explicitly that it refers to $B_H(t)$.
Now, to estimate $H$ for our time series we shall simply adjust the
parameter $H$ so as to obtain the best agreement between the
theoretical curve predicted by $F_H(\tau)$ and our empirical data for
$F(\tau)$. In this way, we can estimate $H$ with a one-parameter
fitting! In principle, one could carry out a formal nonlinear
regression using (\ref{eq:FH}) and (\ref{eq:CH}) to determine $H$, but
in practice it suffices to vary $H$ incrementally and decide by visual
inspection when the theoretical curve best matches the empirical
data. We shall now apply this methodology to estimate the Hurst
exponent for the Ibovespa returns.

\section{Fluctuation analysis of the Ibovespa}

\begin{figure}
\begin{center}
\includegraphics*[width=.7\columnwidth]{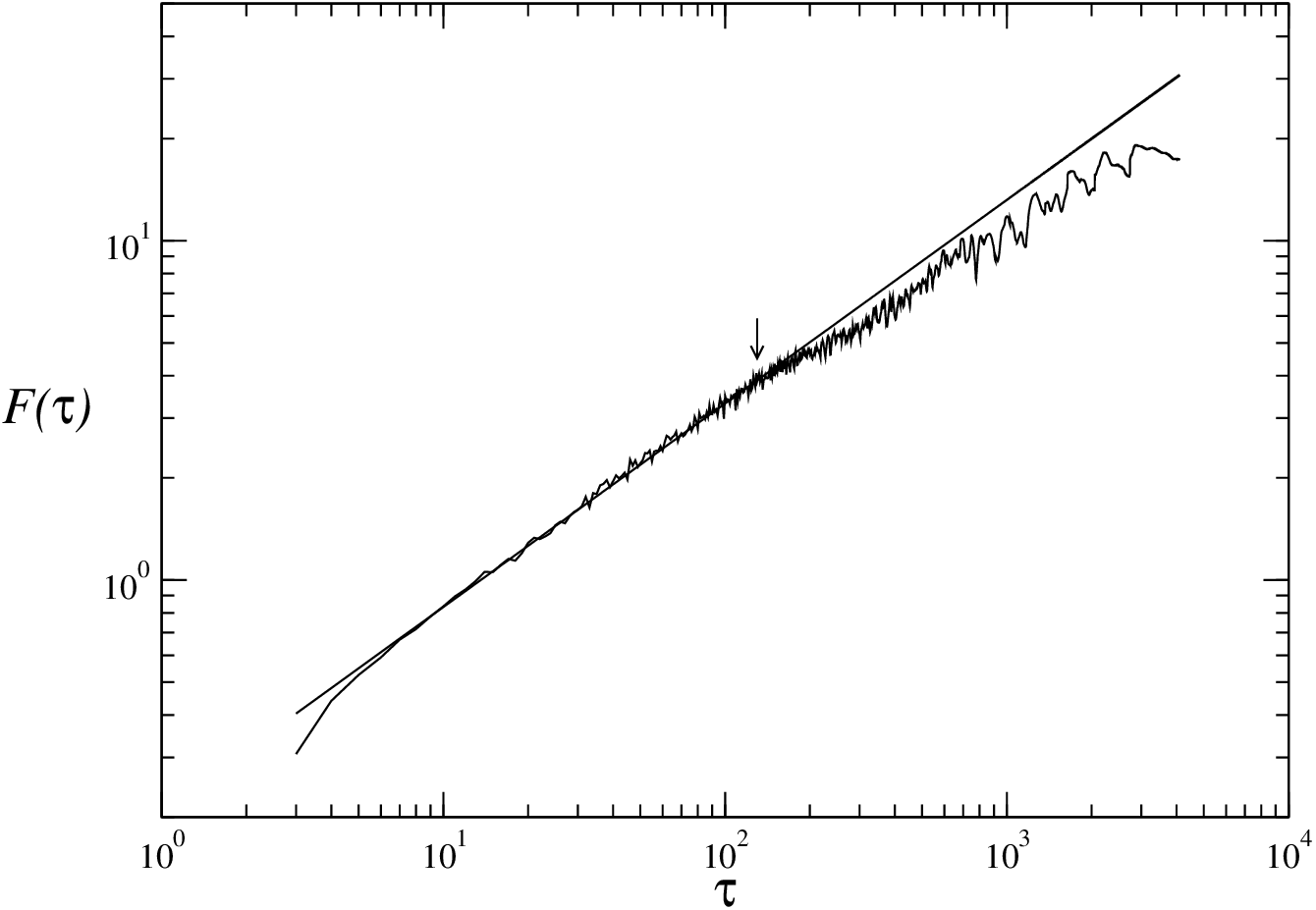}
\caption{Fluctuation function $F(\tau)$ as a function of window size 
$\tau$ for the returns of the Ibovespa index in the period 1968--2001. The
straight line gives the theoretical curve $F_H(\tau)$ for $H=0.6$; see text.}
\label{fig:dfa}
\end{center}
\end{figure}

In this section we discuss the results of the DFA applied to our
Ibovespa time series, which we recall spans the period from January
1968 through May 2001. In Fig.~\ref{fig:dfa} we plot in
double-logarithmic scale the corresponding fluctuation function
$F(\tau)$ against the window size $\tau$.  Using the procedure
outlined at the end of the previous section, we obtain the following
estimate for the Hurst exponent: $H=0.60\pm 0.01$.  In
Fig.~\ref{fig:dfa} we have also plotted (straight line) the
corresponding theoretical curve $F_H(\tau)$ given by (\ref{eq:FH}) and
(\ref{eq:CH}), which as one can see is in excellent agreement with the
empirical data for intermediate values of $\tau$. Since $H>1/2$ we
conclude that the Ibovespa returns show persistence.  We also note
that for $\tau>130$ (indicated by the arrow in Fig.~\ref{fig:dfa}) the
empirical data deviate somewhat from the initial scaling behavior and
appear to cross over to a regime with a slope closer to $1/2$. This
indicates that the Ibovespa tends to loose its `memory' after a period
of about 6 months. It is perhaps worth pointing out that similar
behavior was seen, for example, in several equity indices of the
London  Stock Exchange \cite{london}. There, the break
from the scaling regime typically occurred around 160 trading days
\cite{london}, which is not very different from what we found for the
Ibovespa, especially considering that the location of the crossover
point is difficult to determine precisely.

\begin{figure}
\begin{center}
\includegraphics*[width=.7\columnwidth]{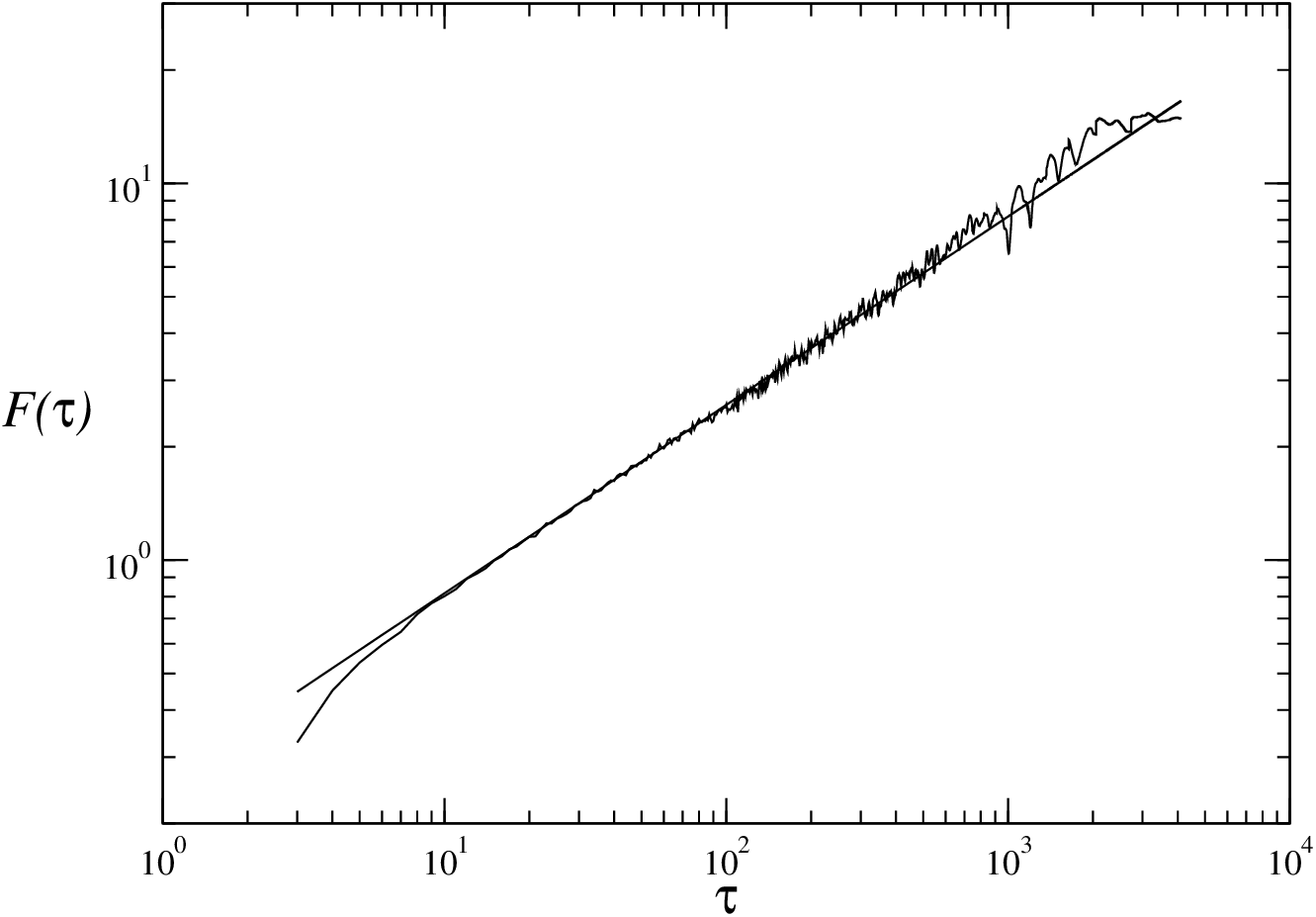}
\caption{Fluctuation function $F(\tau)$  for the shuffled Ibovespa
 returns. The straight line represents the theoretical curve
 $F_H(\tau)$ for $H=0.5$}
\label{fig:random}
\end{center}
\end{figure}

In order to check the validity of our analysis above, we have also
computed the Hurst exponent for a shuffled version of our data in
which we have randomly mixed the time series of the Ibovespa
returns. Because the process of shuffling tends to destroy any
previously existing correlation, we would now expect a Hurst exponent
equal (or very close) to 1/2.  This is indeed the case, as shown in
Fig.~\ref{fig:random}, where we plot $F(\tau)$ for the shuffled
data. In this figure one sees that the data points are extremely well
described by the theoretical curve (straight line) given by
$F_H(\tau)$ with $H=0.5$.  This thus confirms the fact that the
long-term dependence seen in the original time series was not an
spurious effect.

We now wish to investigate whether the Hurst exponent for the Ibovespa
varies in time, which would indicate the existence of nonstationary
fluctuations in the Brazilian stock market.  In fact, a simple visual
inspection of the distribution of the Ibovespa returns shown in
Fig.~\ref{fig:returns} already revels evidences of nonstationary
behavior, which we would like to quantify better. Furthermore, in the
recent past not only the Brazilian economy was plagued by runaway
inflation and but it also endured several `shock-therapy' economic
plans, as we have already mentioned. It is thus natural to ask how
such stressful events may have affected the dynamics of the Ibovespa
index, particularly in regard to its degree of correlation.  We now
turn our attention to this point.

\begin{figure}
\begin{center}
\includegraphics*[width=.7\columnwidth]{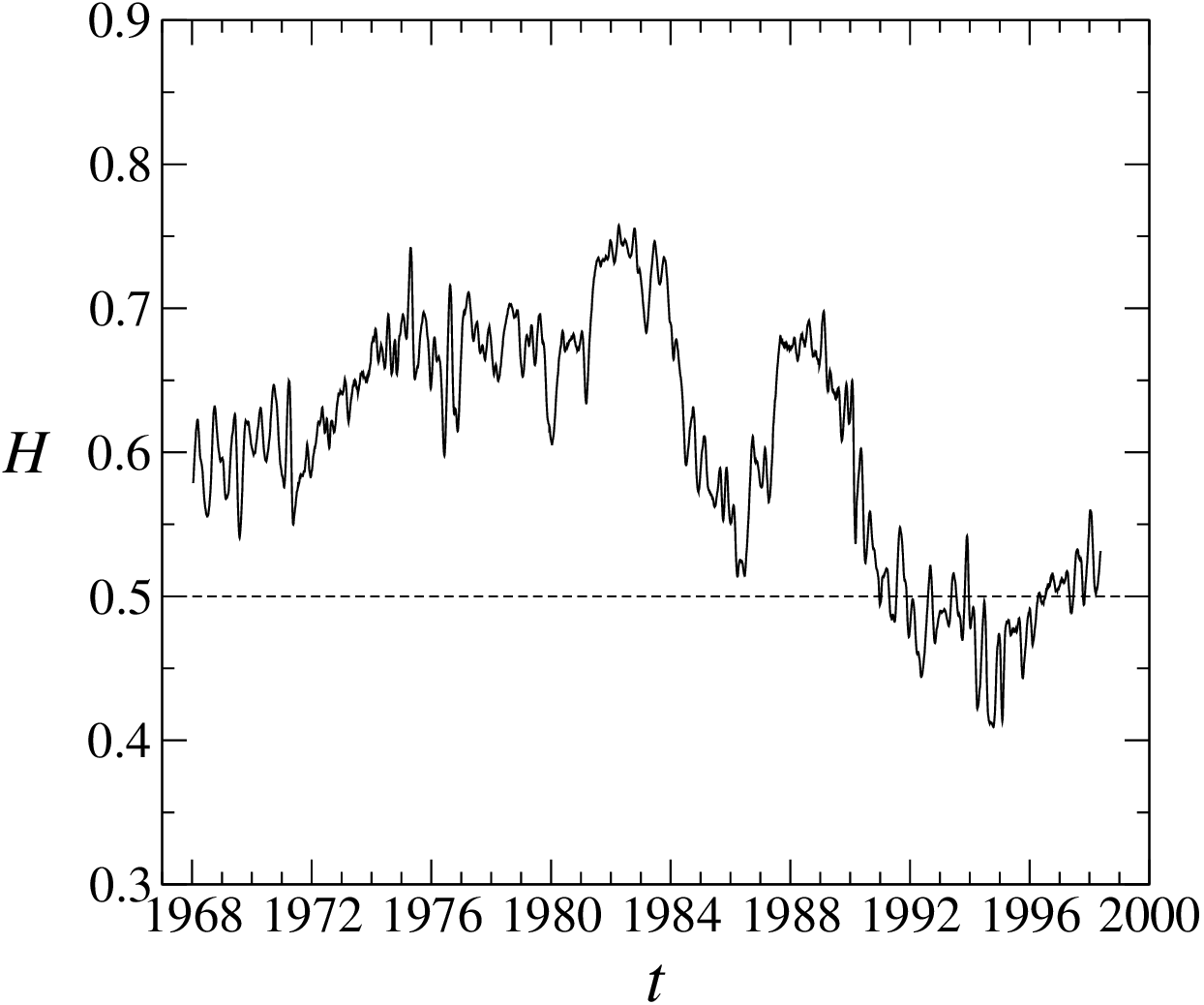}
\caption{The Hurst exponent $H$ for the Ibovespa as a function of time.
Here $H$ was computed in three-year periods and the variable $t$
represents the origin of each such interval.}
\label{fig:bianual}
\end{center}
\end{figure}

In order to estimate a time-varying Hurst exponent using a fluctuation
analysis (or any method for that matter) one is presented with a
difficult challenge, for one has to try to satisfy competing
requirements. Indeed, to determine a `local' exponent $H$ at a
particular time $t$ one has to consider a time interval around $t$
that is considerably smaller than the total time spanned by the data
but still sufficiently large to contain enough points for a meaningful
statistics.  In the case of the Ibovespa we have found that a
three-year period (736 trading days) is an acceptable compromise
between these two opposing demands. (For smaller periods the
fluctuations in the values of $H$ are too large to be trusted.) We
have accordingly applied the DFA to the Ibovespa returns in moving
three-year intervals. More specifically, our methodology is as
follows: starting at the beginning of our time series we compute the
exponent $H$ considering only the data points within three years from
the initial point, then we advance our three-year time window by 20
days and compute $H$ for this new period, and so on. Proceeding in
this way we obtain the curve for $H(t)$ shown in
Fig.~\ref{fig:bianual}, where $t$ denotes the origin of each
three-year time interval.

The most striking feature seen in Fig.~\ref{fig:bianual} is perhaps
the fact that the Ibovespa shows persistence ($H>1/2$) all the way up
to the early 1990's, after which it `switches' to a regime with
alternating persistent and antipersistent behavior but where $H$
remains somewhat close to 1/2. Notice in particular that during the
1970's and 1980's the curve $H(t)$ stays well above 1/2, the only
exception to this trend occurring around the year 1986 when $H$ dips
momentarily towards 1/2---we associate this effect with the launch of
the Cruzado Plan in February 1986; see Sec.~2.  Shortly after the
Cruzado Plan, however, inflation picks up again (Fig.~\ref{fig:igp}),
the Ibovespa declines (Fig.~\ref{fig:ibovespa}), and the exponent $H$
returns to pre-Cruzado levels (Fig.~\ref{fig:bianual}). Then in the
early 1990's, at the time of the Collor Plan (see below), we observe a
dramatic decline in the curve $H(t)$ towards 1/2, with $H$ remaining
(within some fluctuation) around this value afterwards. Note also that
around the time of the launch of the Real Plan in July 1994 the curve
$H(t)$ reaches its lowest values. We have thus seen that the Ibovespa
Hurst exponent declines following the adoption of a major economic
plan (such as the Cruzado, Collor and Real Plans), thus confirming the
fact \cite{ausloos} that a firm intervention on the market tends to
reduce $H$ momentarily.

\begin{figure}
\begin{center}
\includegraphics*[width=.7\columnwidth]{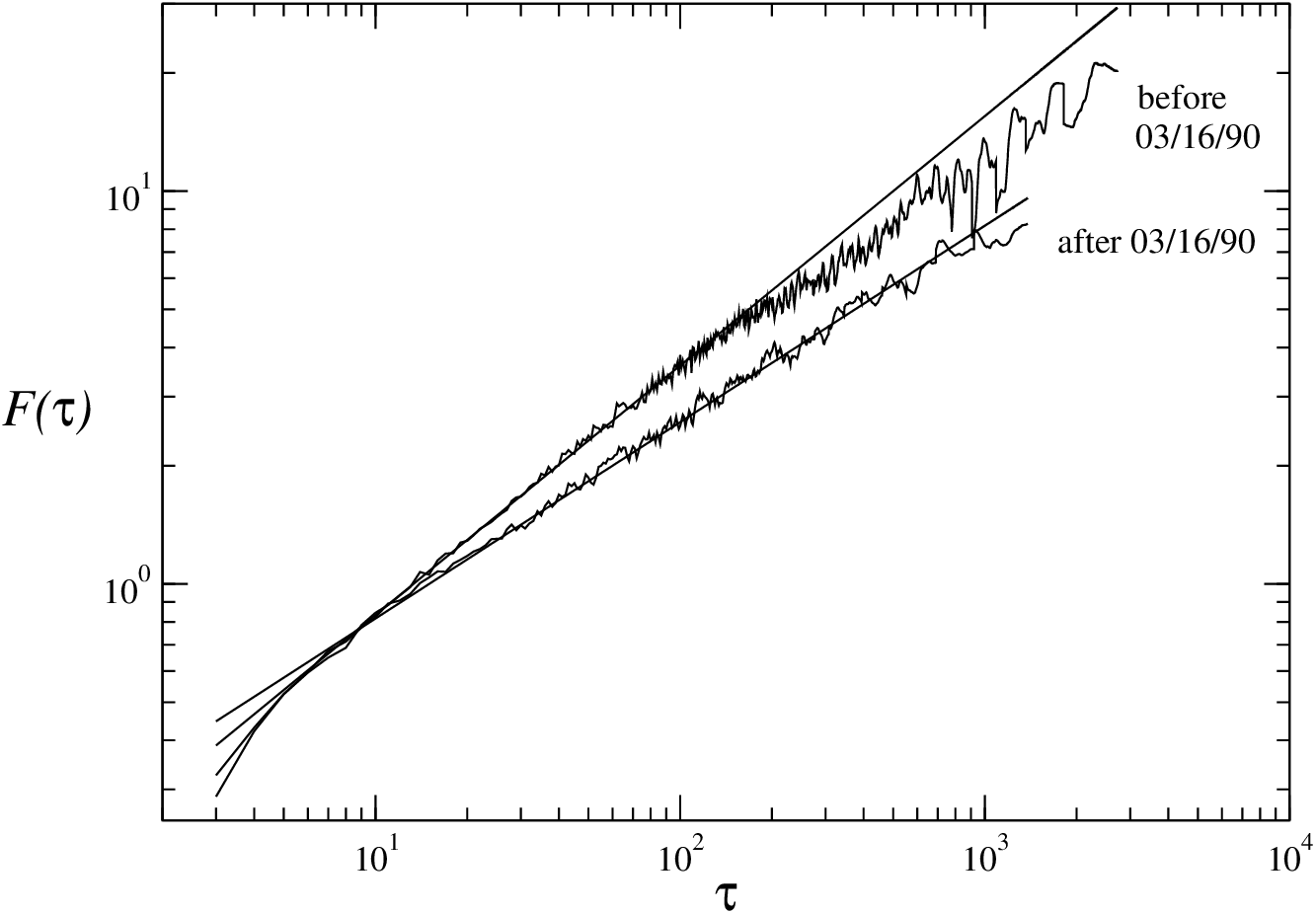}
\caption{Plot of $F(\tau)$  for the  Ibovespa returns 
before and after the Collor Plan. The upper (lower) straight line
corresponds to the theoretical curve $F_H(\tau)$ with $H=0.63$
($H=0.5)$.}
\label{fig:collor}
\end{center}
\end{figure}

As we already mentioned, the drop in $H$ that occurs in the early
1990's can be unambiguously traced back to the launch, in March 1990,
of the Collor Plan which marked the beginning of structural reforms in
Brazil \cite{FMI}.  In this context, it is important to realize that
this decline in $H$ takes place well before the adoption of the Real
Plan.  It is indeed quite surprising that the Hurst exponents found in
the early 1990's, in a macroeconomic environment dominated by very
high inflation, are on about the same level as those found after the
Real Plan when the economy had been stabilized.  This fact in itself
is a clear evidence of the profound impact that the structural reforms
set off by the Collor Plan had on the Brazilian economy.  In order to
study this effect in more detail, we have applied the DFA separately
to the Ibovespa returns prior and after the Collor Plan. Our results
are shown in Fig.~\ref{fig:collor}. In this figure one sees that
before the Collor Plan the Brazilian stock market shows a significant
amount of persistence ($H=0.63$), while after the Collor Plan the
Ibovespa displays essentially no long-term dependence ($H=0.5$).

In light of the preceding results, it thus seems legitimate to
conclude that the opening and consequent modernization \cite{FMI} of the
Brazilian economy that begun in the early 1990's  has led to
a more efficient stock market, in the sense that $H=0.5$ for the
Ibovespa after 1990. By the same token, we associate the period of
higher $H$ before 1990 with a less efficient market. In those years,
the Brazilian economy was considerably closed to foreign competition
and its financial markets were not readily accessible to international
investors. In such closed economic environment the Brazilian stock market was
conceivably more prone to `correlated fluctuations' (and perhaps also
more susceptible to being pushed around by aggressive investors), which
may explain in part a Hurst exponent greater than 0.5 for that period.

\section{Multifractional Brownian motion}

We have seen above that the Hurst exponent $H$ for the Ibovespa stock
index changes appreciably over time. In this regard, the fractional
Brownian motion appears to be a somewhat restrictive model, unable to
capture more fully the complex dynamics of the Brazilian stock market.
Similar time-dependent Hurst exponents have also been observed in
other financial markets \cite{ausloos,ausloos2,malaio} and in physical
processes such as traffic flow in the Internet \cite{traffic}.  To
circumvent this shortcoming of the FBM, Peltier and Levy-Vehel
\cite{mfbm} have proposed a multifractional Brownian motion (MFBM) in
which the scaling exponent $H$ is allowed to vary in time. For a
rigorous exposition of the definition and mathematical properties of
MFBM the interested reader is referred to Ref.~\cite{mfbm}.  Here we
shall present only the main properties of the MFBM that are of
interest to us. Before we proceed, it should however be emphasized
that our aim in this section is not to develop a specific model for
the Ibovespa but rather to present a generic theoretical framework
based on the MFBM in which the time-dependence of the Ibovespa Hurst
exponent could, in principle, be understood.

Let $H(t)$ be a H\"older-continuous function in the interval
$t\in[0,1]$ with H\"older exponent  $\beta>0$, such that for any $t>0$ we have
$0<H(t)<\rm{min}(1,\beta)$ \cite{mfbm}. (For ease of notation, we
shall sometimes indicate the $t$-dependence as a subscript and write
$H_t$.)  The multifractional Brownian motion
$\{W_{H_t}(t), t>0\}$ is the  Gaussian process defined by
\begin{equation}
W_{H_t}(t)= \frac{1}{\Gamma(H_t+\frac{1}{2})}\int_{-\infty}^t[
(t-s)_+^{H_t-1/2}-(-s)_+^{H_t-1/2}]dB(t),
\end{equation}
where $\Gamma(x)$ is the  gamma function, $(x)_+$ equals $x$ if
$x>0$ and zero otherwise, and $B(t)$ denotes the usual Brownian
motion.  One important aspect of the process defined above is that its
increments are no longer stationary since it can be shown \cite{mfbm} that
\begin{eqnarray}
E[\{W_{H_{t+h}}(t+h)-W_{H_t}(t)\}^2]\approx h^{2H(t)} \quad \mbox{as} \quad h\to 0.
\end{eqnarray}
Because of its non-stationarity the MFBM is no longer a self-similar
process either. However, it is possible to define the concept of
locally asymptotically similarity at the point $t_0>0$ in the
following way
\begin{equation}
 \frac{W_{H_{t_0+at}}(t_0+at)-W_{H_{t_0}}(t_0)}{a^{H_{t_0}}}
 \stackrel{d}{=}W_{H_{t_0}}(t) \quad \mbox{as} \quad a\to0.
\end{equation}
[Compare this with the global self-similar property of the FBM given
in (\ref{eq:aH}).] In this sense, $W_{H_t}(t)$ can be thought of as a
process that locally at time $t$ `resembles' a fractional Brownian
motion $B_{H_t}(t)$. In fact, a practical way to generate a
multifractional Brownian path consists in generating for each time
$t\in[0,1]$ a fractional Brownian motion $B_{H_t}(s)$, $s\in [0,1]$,
and then setting $W_{H_t}(t)=B_{H_t}(t)$. More details about this
method together with an algorithm for its implementation can be found
in Ref.~\cite{mfbm}.  (We remark parenthetically, however, that the
algorithm presented there for the computation of $H(t)$ is not
appropriate for empirical time series, as inadvertently used in
Ref.~\cite{malaio}, because such algorithm requires previous knowledge
of the variance of the FBM used to generate the MFBM, which of course
is not known for empirical data.)

\begin{figure}[t]
\begin{center}
\includegraphics*[width=.47\columnwidth]{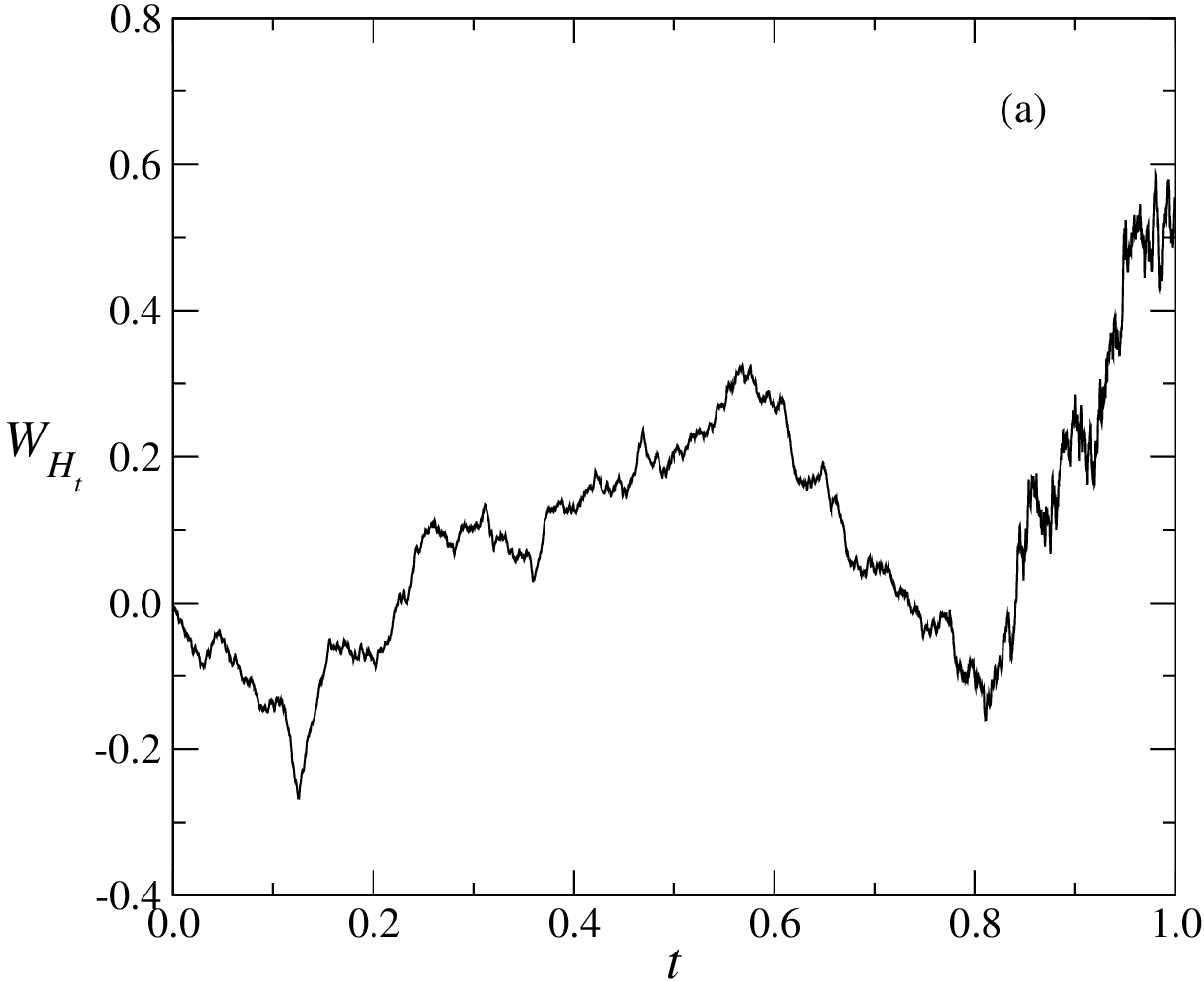}
\quad
\includegraphics*[width=.47\columnwidth]{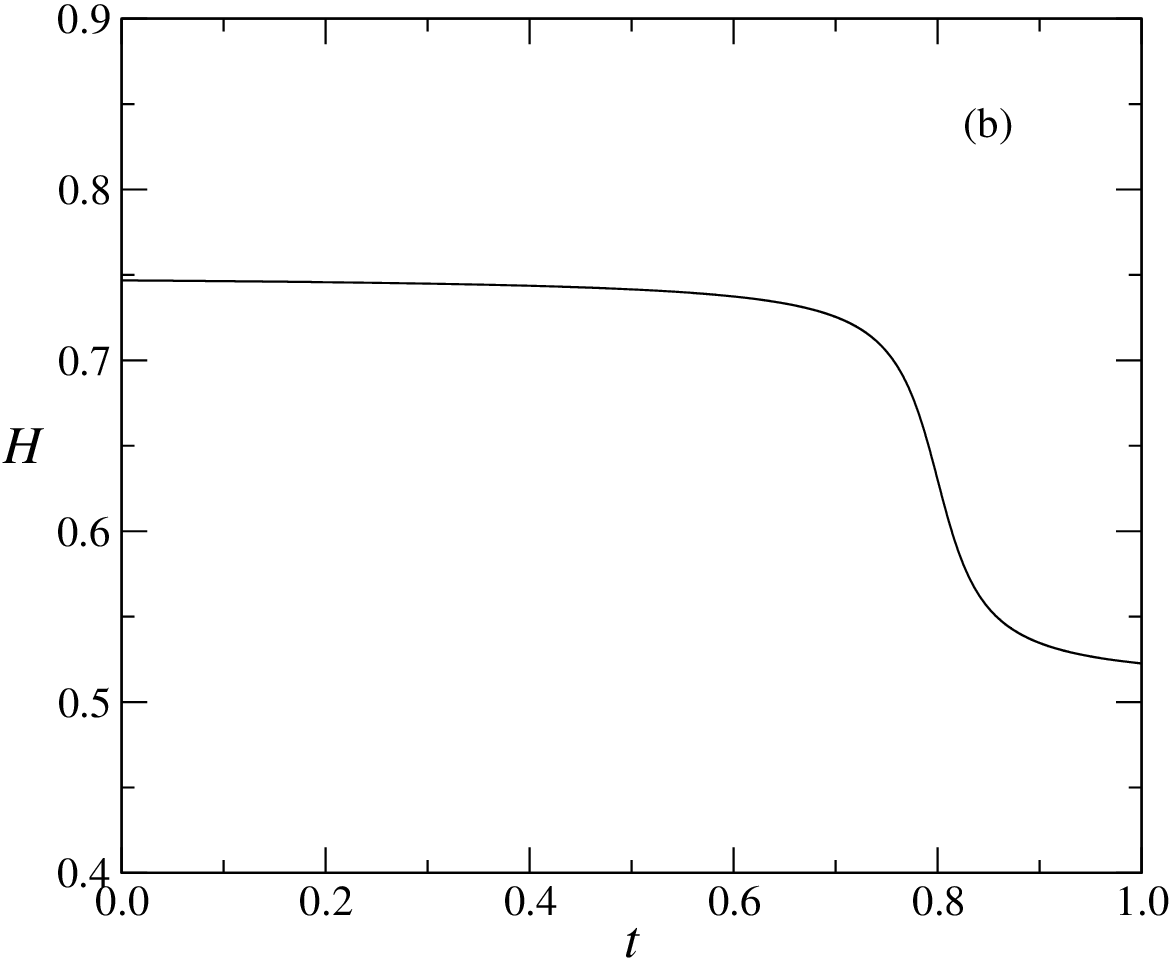}
\caption{Path of a multifractional Brownian motion (a) with 
the H\"older function $H(t) = 0.63 - 0.076\arctan (30t-24)$ shown in
(b).}
\label{fig:mfbm}
\end{center}
\end{figure}

\begin{figure}[t]
\begin{center} 
\includegraphics*[width=.7\columnwidth]{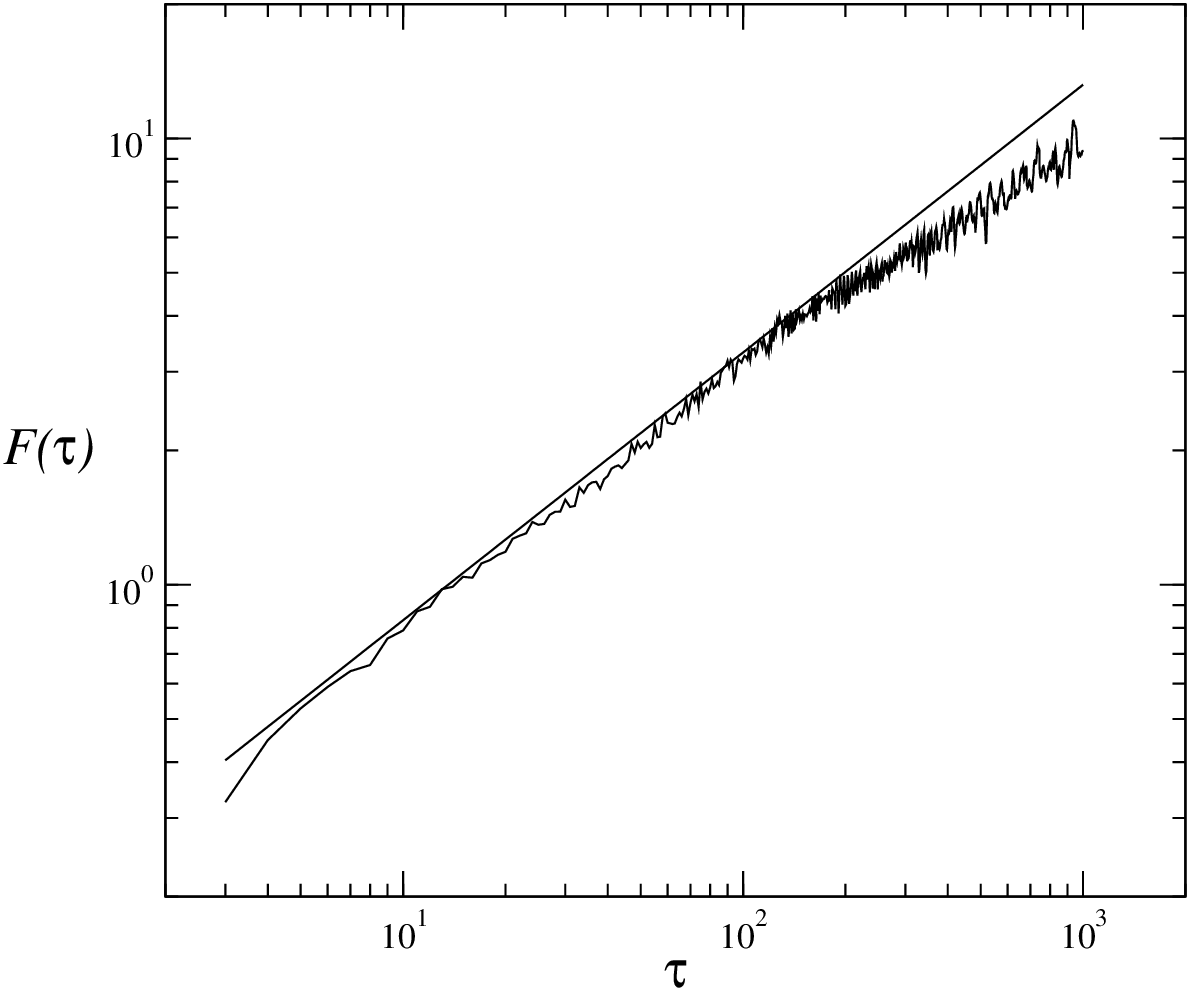} 
\caption{Detrended fluctuation analysis of the multifractional Brownian motion
shown in Fig.~\ref{fig:mfbm}(b).}
\label{fig:dfa_mfbm}
\end{center}
\end{figure}

In Fig.~\ref{fig:mfbm} we show an example of a path of
multifractional Brownian motion characterized by the Hurst function
\begin{equation}
H(t) = 0.63 - 0.076\arctan (30t-24).
\label{eq:atan}
\end{equation}
Although somewhat arbitrary, this example was chosen to mimic the
generic trend seen in the Ibovespa where $H(t)$ can be viewed as
having (on average) two distinct `plateaus,' one before and the other
after 1990, with a swift crossover between them.  For comparison
purpose, we have generated a MFBM path with approximately the same
number of data points ($N=2^{13}=8192$) as in the original Ibovespa
time series ($N=8209$). We have then applied the detrended fluctuation
analysis to our MFBM path, with the result being shown in
Fig.~\ref{fig:dfa_mfbm}. Since a MFBM path is not characterized by a
constant Hurst exponent, the functional dependence of the DFA
fluctuation function $F(\tau)$ will not, in general, be a power
law. Nevertheless, we see from Fig.~\ref{fig:dfa_mfbm} that $F(\tau)$
exhibits an approximate scaling regime, $F(\tau)\sim\tau^{\bar{H}}$
with $\bar{H}=0.6$, up to about $\tau=160$.  For comparison, we also show
in Fig.~\ref{fig:dfa_mfbm} the corresponding relationship (straight
line) predicted by $F_H(\tau)$ given in (\ref{eq:FH}), which as one
sees is in good agreement with the data in the scaling region.

We thus interpret the scaling exponent $\bar{H}$ as a kind of average
Hurst exponent for our MFBM time series. Note, however, that although
the region of small $H$ represents only about 20\% of the total time
series it has nonetheless a significant weight on the exponent
$\bar{H}$, since its  value  is considerably
smaller than what one would obtain by simply averaging the curve
$H(t)$ [in this case one gets $H_{\rm ave}=0.68$]. This result is in
agreement with the known fact \cite{stanley2} that regions of large
fluctuations (and hence smaller $H$) tend to dominate the behavior of
$F(\tau)$ for small-to-intermediate values of $\tau$.  A more detailed
discussion about the detrended fluctuation analysis of MFBM paths will be
presented elsewhere.

The preceding discussion thus suggests that the Hurst exponent $H$
obtained by applying the DFA (and similar fluctuation analyses
\cite{taqqu}) to financial time series might perhaps capture
only an average behavior over the time period spanned the data. This
seems to be particularly true for the Ibovespa index, where we find
that the DFA Hurst exponent obtained using the complete time series
(spanning over 30 years of data) is $H=0.6$, whereas during this
period the local exponent $H(t)$ varies considerably within the range
$(0.76,0.42)$; see Fig.~\ref{fig:bianual}.  One is thus lead to
conclude that when dealing with long financial time series, a more
localized (in time) analysis is necessary to determine a possible
dependence of $H$ on time. In particular, note that a global Hurst exponent
equal to 1/2 does not necessarily imply the absence of correlation:
there could be positive and negative correlations at different periods
of time that average out so as to yield an effective $H=0.5$.

\section{Conclusions}

We have carried out a Detrended Fluctuation Analysis of the daily
returns of the S\~ao Paulo Stock Exchange Index (Ibovespa), spanning
over 30 years of data from January 1968 up to May 2001. For this time
series we have obtained a Hurst exponent greater than 1/2, indicating
that the Ibovespa has long-term dependence (persistence). This memory
effect seems to last for up to about 6 months (130 trading days),
after which time the fluctuation function $F(\tau)$ deviates from the
initial scaling behavior and  crosses over to a regime with a
slope closer to 1/2. Similar behavior has been observed in other stock
indices \cite{london}, and it thus appears that stock markets tend to
loose their `memory' typically in about half a year.

We have also performed a more localized (in time) analysis of the
Ibovespa returns, where we have calculated the local Hurst exponent
$H(t)$ in three-year moving time windows. Here the most striking feature is
the fact that after 1990 the Hurst exponent stays close to 1/2 (within
some fluctuation), whereas in the preceding decades it was, in general,
considerably larger than 1/2. We have thus argued that the
liberalizing measures (e.g., substantial reduction of import tariffs,
cutting down the `red tape' for foreign investments, and privatisation of
state companies \cite{FMI}) brought about by the Collor Plan in the
early 1990's and continued with the Real Plan in 1994 resulted in a more
efficient stock market in the sense that $H$ remained close to 1/2
after 1990.

The fact that its Hurst exponent changes over time time indicates that
the Ibovespa time series cannot be satisfactorily modeled in terms of
a simple fractional Brownian motion. In other words, the Ibovespa
follows a multifractal process. We have thus suggested that the
multifractional Brownian motion, in which the scaling exponent
$H$ is allowed to vary in time, could perhaps be a more appropriate
description for the Brazilian stock market dynamics. It thus remains
an interesting question to investigate to what extent the MFBM is a
valid model for other financial time series.

\begin{ack}
We thank A.~Ohashi for fruitful discussions at early stages of this
work.  Stimulating discussions on related topics with D.~Rothman and
R.~Almgren are also acknowledged. This research was supported in
part by the Brazilian agencies FINEP and CNPq and the special program
PRONEX.
\end{ack}

\appendix
\section{Appendix: Hurst rescaled range analysis}

The Hurst rescaled range ($R/S$) analysis is a technique devised by
the hydrologist Henry Hurst in 1951 \cite{hurst} to test for the
presence of correlations in empirical time series. A detailed
description of the Hurst analysis (including its historical origin)
can now be found in textbooks \cite{feder,turcotte}, so that here we
shall give only a brief summary of the method and then proceed
to  apply it to the Ibovespa returns.

In the Hurst analysis, we start by dividing our time series
$\{r_t\}_{t=1,...,T}$ in $N$ non-overlapping time intervals $I_n$ of
equal size $\tau$. For each interval $I_n$ we  compute the range
\begin{equation}
R_n =\max _{1\leq k\leq \tau } \left[ \sum _{t=1}^{k}\left(
r_{n\tau+t}-\overline{r}_n \right) \right] - \min _{1\leq
k\leq \tau }\left[ \sum _{t=1}^{k}\left(
r_{n\tau+t}-\overline{r}_n\right) \right],
\end{equation}
where
\begin{equation}
\overline{r}_n=\frac{1}{\tau}\sum_{t=1}^{\tau}  r_{n\tau+t},
\end{equation}
and the data
standard deviation
\begin{equation}
S_n=\sqrt{\frac{1}{\tau }\sum _{t=1}^{\tau }\left(
r_{n\tau+t}-\overline{r}_n\right) ^{2}} .
\end{equation}
The rescaled range $(R/S)_\tau$ is  then defined as the average of the ratio
$R_n/S_n$ over all intervals $I_n$
\begin{equation}
\left(R/S\right)_\tau \equiv \left\langle \frac{R_n}{S_n}\right\rangle = \frac{1}{N} \sum_{n=0}^{N-1}  \frac{R_n}{S_n}.
\end{equation}
The Hurst exponent $H$ is obtained from the scaling behavior of 
$(R/S)_\tau$:
\begin{equation}
\left(R/S\right)_\tau \sim \tau^{H}.
\label{eq:H}
\end{equation} 

\begin{figure}[t]
\begin{center}
\includegraphics*[width=.7\columnwidth]{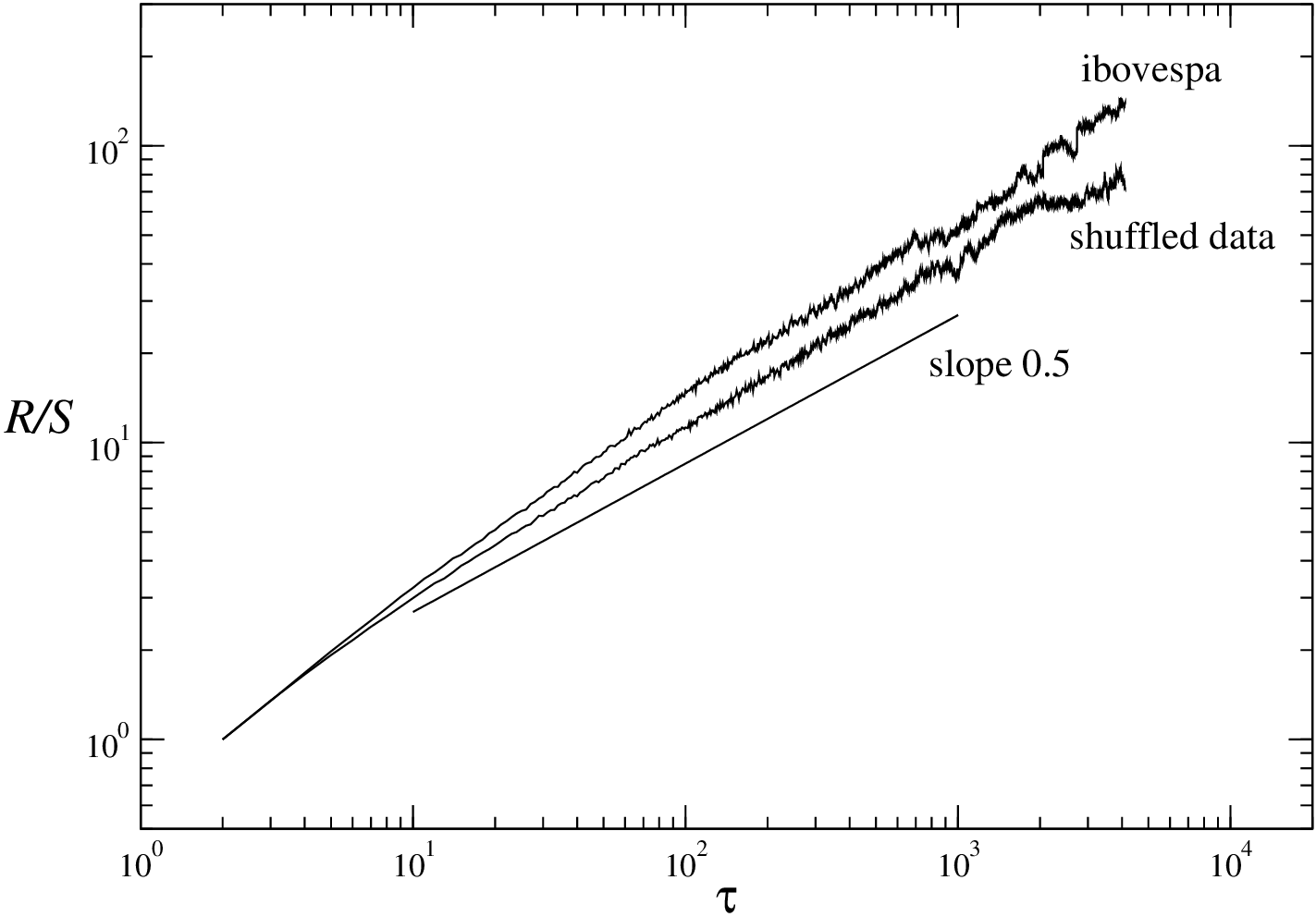}
\caption{Rescaled range $R/S$ versus the time lag $\tau$ for
the Ibovespa returns in the period 1968---2001 and for the corresponding
shuffled data. Also shown for comparison is a straight line with slope
equal to 1/2.}
\label{fig:hurst}
\end{center}
\end{figure}

A plot of the rescaled range $R/S$ as a function of $\tau$ for the
Ibovespa returns over the entire period (1968--2001) analyzed in this
paper is shown in the upper curve of Fig.~\ref{fig:hurst}. The data in
this case show a scaling regime that goes from $\tau=10$ up to about
$\tau=160$. A linear regression in this region yields the value
$H=0.65$, which is somewhat larger than the exponent ($H=0.6$)
obtained via the DFA for the same time series. In order to understand
the origin of such discrepancy, we have also applied the Hurst method
to the shuffled Ibovespa data (lower graph in Fig.~\ref{fig:hurst})
and found $H=0.57$ in this case. We thus see that the Hurst analysis
of the shuffled data predicts an unexpected residual correlation.
(Recall that applying the DFA to the shuffled data yields $H=0.5$, as
expected.) This `excess of correlation' found in the $R/S$ analysis of
the shuffled Ibovespa is perhaps not entirely surprising, given that
it is known that the Hurst method tends to overestimate the Hurst
exponent for time series of small sizes \cite{feder}.  This effect may
perhaps explain why the exponent $H$ obtained via the Hurst method is
usually larger than that found via the DFA \cite{pilar}. We thus
conclude that the DFA appears to be a more reliable method for
detecting true correlation in financial time series. It is worth
therefore emphasizing that the numerical values obtained for the
exponent $H$ via the Hurst method should be interpreted with some
caution, specially when the time series is not sufficiently large,
as is commonly the case in practice.

\end{document}